\documentclass[conference]{IEEEtran}
\usepackage[T1]{fontenc}
\usepackage{float}
\usepackage{dblfloatfix}
\usepackage[dvips]{graphicx}
\graphicspath{{../eps/}}
\DeclareGraphicsExtensions{.eps}
\ifCLASSINFOpdf
\else
\fi

\ifCLASSOPTIONcompsoc
  \usepackage[caption=false,font=normalsize,labelfont=sf,textfont=sf]{subfig}
\else
  \usepackage[caption=false,font=footnotesize]{subfig}
\fi

\usepackage{amsmath}
\usepackage[cmintegrals]{newtxmath}
\usepackage{graphicx}
\newtheorem{theorem}{Theorem}

\newtheorem{lemma}{Lemma}
\begin{document}

\title{An Outage Probability Analysis of Full-Duplex NOMA in UAV Communications}

\author{
\IEEEauthorblockN{ Tan Zheng Hui Ernest*, A S Madhukumar*, Rajendra Prasad Sirigina*, Anoop Kumar Krishna**}
\IEEEauthorblockA{*School of Computer Science and Engineering\\
Nanyang Technological University, Singapore\\
Email: tanz0119@e.ntu.edu.sg, \{raje0015, asmadhukumar\}@ntu.edu.sg}
\IEEEauthorblockA{**Airbus Singapore Pte Ltd, Singapore\\
Email: anoopkumar.krishna@airbus.com}
}

\maketitle

\begin{abstract}
As unmanned aerial vehicles (UAVs) are expected to play a significant role in fifth generation (5G) networks, addressing spectrum scarcity in UAV communications remains a pressing issue. In this regard, the feasibility of full-duplex non-orthogonal multiple access (FD-NOMA) UAV communications to improve spectrum utilization is investigated in this paper. Specifically, closed-form outage probability expressions are presented for FD-NOMA, half-duplex non-orthogonal multiple access (HD-NOMA), and half-duplex orthogonal multiple access (HD-OMA) schemes over Rician shadowed fading channels. Extensive analysis revealed that the bottleneck of performance in FD-NOMA is at the downlink UAVs. Also, FD-NOMA exhibits lower outage probability at the ground station (GS) and downlink UAVs than HD-NOMA and HD-OMA under low transmit power regimes. At high transmit power regimes, FD-NOMA is limited by residual SI and inter-UAV interference at the downlink UAVs and FD-GS, respectively. The impact of shadowing is also shown to affect the reliability of FD-NOMA and HD-OMA at the downlink UAVs.
\end{abstract}

\begin{IEEEkeywords}
Unmanned Aerial Vehicle, Full-Duplex, NOMA, Outage Probability, Rician Shadowed Fading.
\end{IEEEkeywords}

\IEEEpeerreviewmaketitle

\section{Introduction}

With the deployment of fifth generation (5G) networks likely to occur in the near future, the role of unmanned aerial vehicles (UAVs) in 5G networks is expected be prominent. Already, diverse applications involving multi-UAV use cases have been envisaged, such as UAV-based cellular communications in disaster areas \cite{yadav2018full} and geographical surveying \cite{andre2014application}.

However, spectrum scarcity in UAV communications remains a pressing issue, despite the adoption of 5G-related technologies such as power-domain non-orthogonal multiple access (NOMA). In the literature, nodes in conventional NOMA and orthogonal multiple access (OMA) schemes are assumed to operate in half-duplex (HD) mode. As a result, orthogonal spectrum allocation for uplink and downlink users is required.
With the allocated bands for UAV communications also shared by many other existing systems \cite{matolak2017air_suburban}, spectrum scarcity is not adequately addressed when satisfying such a requirement.

In this regard, the full-duplex NOMA (FD-NOMA) paradigm is worth considering as an attractive and pragmatic alternative \cite{ding2018coexistence}. When applied in the context of UAV communications, FD-NOMA enables uplink UAVs, downlink UAVs, and FD ground stations (FD-GSs) to simultaneously communicate on the same spectrum, with the feasibility of FD-NOMA having been investigated in \cite{ding2018coexistence} from a sum rate perspective.

Although FD-NOMA enables spectrum efficiency to be improved, performance limiting factors are introduced as a consequence. Specifically, self-interference (SI) and inter-UAV interference are experienced at the FD-GS and the downlink UAVs, respectively. SI at the FD-GS can be suppressed through SI mitigation architectures operating in the digital or analog domain \cite{sahai2013impact}. However, hardware impairments can cause SI mitigation to be imperfect due to phase noise and imperfect SI channel estimation \cite{sahai2013impact}. As a result, the remaining residual SI is a performance limiting factor at the FD-GS. At the downlink UAVs, inter-UAV interference from uplink UAVs and power-domain NOMA reduces the effectiveness of FD-NOMA. Thus, effective interference management strategies must be employed at the downlink UAVs. 

Apart from performance limiting factors, FD-NOMA UAV communications is also impaired by small scale fading and shadowing, especially in the suburban environment due to the occurrence of line-of-sight blockage, i.e., Rician shadowed fading, \cite{tan2018ricianShad}. Although Rician shadowed fading has been studied in UAV communications to model the combined effect of fading and shadowing \cite{tan2018ricianShad}, similar studies in the context of FD-NOMA UAV communications over Rician shadowed fading channels is still lacking.

To this end, the feasibility of FD-NOMA in the context of UAV communications is investigated in this paper. In particular, the reliability of FD-NOMA UAV communications is compared against conventional NOMA and OMA schemes, i.e., HD-NOMA and HD-OMA. The major contributions of this paper are as follows.
\begin{itemize}
	\item Closed-form outage probability expressions for FD-NOMA, HD-NOMA, and HD-OMA schemes in UAV communications over Rician shadowed fading channels are presented.
	\item At low transmit power regimes, it is shown that FD-NOMA exhibits lower outage probability than HD-NOMA and HD-OMA at the GS and at the downlink UAVs. However, FD-NOMA is limited by residual SI at the GS and inter-UAV interference at the downlink UAVs at high transmit power regimes. At the system level, it is observed that the reliability of FD-NOMA is limited by the downlink UAVs.
	\item It is also demonstrated that the presence of shadowing adversely affects outage probability at the downlink UAVs for FD-NOMA, HD-NOMA, and HD-OMA schemes. 
\end{itemize}

The organization of this paper is as follows. The system model is introduced in Section \ref{sec_sys_model}, with outage probability expressions presented in Section \ref{sec_outage}. Numerical results are discussed in Section \ref{sec_num_res} before the conclusion of the paper in Section \ref{sec_conclusion}.

\section{System Model} \label{sec_sys_model}
\begin{figure} [tpb]
\centering
\includegraphics [width=0.95\columnwidth]{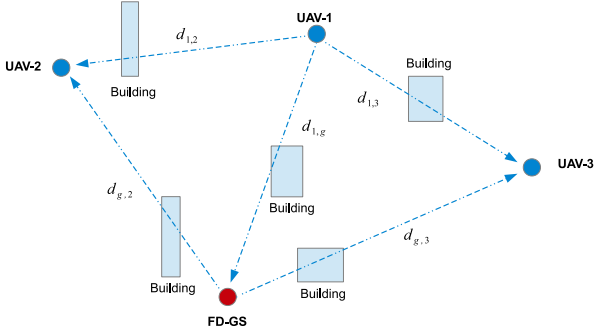} 
\vspace{-1cm}
\caption{An illustration of FD-NOMA UAV communications in a suburban environment. The uplink UAV (UAV-1) and downlink UAVs (UAV-2 and UAV-3) simultaneously communicate on the same spectrum with the FD-GS.}
\label{fig:block_diagram}
\end{figure}

An FD-NOMA unmanned aerial vehicle (UAV) network consisting of half-duplex (HD) UAVs and an FD-GS in a suburban environment is considered in this work (Fig. \ref{fig:block_diagram}). In the particular scenario depicted in Fig. \ref{fig:block_diagram}, the FD-GS receives uplink data from Unmanned Aerial Vehicle 1 (UAV-1). Simultaneously, Unmanned Aerial Vehicle 2 (UAV-2) and Unmanned Aerial Vehicle 3 (UAV-3) receives downlink data from the FD-GS through power-domain NOMA. Thus, SI and inter-UAV interference is experienced at the FD-GS and the downlink UAVs, respectively. 

To account for the spatial locations of the UAVs, let the Euclidean distance (in km) between UAV-1 and the FD-GS be $d_{1,g}$. Also, let the inter-UAV Euclidean distance between UAV-1 and UAV-$i$ be $d_{1,i}, i \in \{2,3\}$. Similarly, for the downlink UAVs, let the Euclidean distance between UAV-1 and UAV-$i$ be $d_{g,i}, i \in \{2,3\}$ where $0<d_{g,2}<d_{g,3}$. As it is likely for UAVs to be subjected to altitude restrictions in urban environments, all UAVs are assumed to be operating at the same altitude in this work. Also, Rician shadowed fading channels are assumed for the UAV and the SI channels to model the suburban environment \cite{tan2018ricianShad}.

At the FD-GS, let the transmitted signals from the FD-GS and UAV-1 be $x_{gs}=\sqrt{a_{gs,2}}x_{gs,2}+\sqrt{a_{gs,3}}x_{gs,3}$ and $x_1$, respectively, where $x_{gs,i}$ is the intended message for UAV-$i$ and $a_{gs,i}$ is the power allocation factor at UAV-$i$ such that $a_{gs,2}+a_{gs,3}=1$. Additionally, let $x_1$, $x_{si}=x_{gs}$, $h_{1,g}$, and $h_{si}$ denote the signal-of-interest (SOI), SI signal, channel between UAV-1 and GS, and the SI channel gain at the FD-GS, respectively. Then, the received signal at GS can be written as \cite{sahai2013impact}:
\begin{eqnarray} \label{y_gs}
y_{gs} \hspace{-0.3cm}& = & \hspace{-0.3cm} \sqrt{\frac{P_t}{d_{1,g}^{n}}}h_{1,g}x_{1} \hspace{-0.05cm} + \hspace{-0.05cm} \sqrt{P_{si}} |\widetilde{h}_{si}|x_{si} \hspace{-0.05cm} + \hspace{-0.05cm} \sqrt{P_{si}}|h_{si}|\gamma_{\phi}w_{\phi} \hspace{-0.05cm} + \hspace{-0.05cm} w_{g},
\end{eqnarray}
where $n$ is the pathloss exponent, $P_t$ is the transmit power normalized by receiver noise, $P_{si}$ is the power of the SI, $\widetilde{h}_{si}$ is the error of the imperfect SI channel gain estimate, defined as $\widetilde{h}_{si}=h_{si}-\widehat{h}_{si}$, $\widehat{h}_{si}$ is the imperfect estimation of the SI channel gain, $w_{g}$ is the additive white Gaussian noise (AWGN) with zero-mean and variance $\sigma_g^2$, and $w_{\phi}$ is the Gaussian distributed phase noise with zero-mean and unit variance scaled by the strength of the phase noise $\gamma_{\phi}^2$ \footnote{The phase noise term $\gamma_{\phi}$ reflects the jitter effect in oscillators due to hardware imperfections \cite{sahai2013impact}} \cite{sahai2013impact}. To model the worst case residual SI, the channel estimation error ($\widetilde{h}_{si}$) is modeled as a circularly symmetric zero-mean complex Gaussian random variable (RV) with unit variance scaled by $\epsilon$ \cite{zlatanov2017capacity}.

At the downlink UAVs, the received signal at UAV-$i$ for $i \in \{2,3\}$ can be written as:
\begin{eqnarray}
y_{i} & = & \sqrt{\frac{P_t}{d_{g,i}^{n}}}h_{g,i}x_{gs} + \sqrt{\frac{P_t}{d_{1,i}^{n}}}h_{1,i}x_{1} + w_{i},
\end{eqnarray}
where $h_{g,i}$ denotes the channel between the FD-GS and UAV-$i$, $h_{1,i}$ is the inter-UAV interference channel between the UAV-1 and UAV-$i$, and $w_{i}$ is the AWGN at UAV-$i$ with zero-mean and variance $\sigma_i^2$.

As FD-NOMA is considered in this work, inter-UAV interference is present at the downlink UAVs due to power-domain NOMA and uplink interference. Thus, an imperfect SIC detector is assumed at UAV-2 while an interference ignorant (II) detector is assumed at UAV-3. The SIC detector at UAV-2 removes $x_{gs,3}$ before detecting $x_{gs,2}$ while treating $x_1$ as noise. At UAV-2, the II detector treats both $x_{gs,2}$ and $x_{1}$ as noise while detecting $x_{gs,3}$.

\section{Outage Probability} \label{sec_outage}

The FD-NOMA outage probability expressions at the FD-GS and the downlink UAVs are presented in this section. For benchmark comparison, the outage probability expressions for HD-NOMA and HD-OMA are also presented. Let the respective transmission rates of UAV-1 and GS be $R^{i}_{1}$ and $R^{i}_{gs}$ for $i \in \{FD, HD, HD-OMA\}$, with sum rate defined as $R^{i}_{sum} = R^{i}_{1}+R^{i}_{gs}$. To ensure a fair comparison between FD-NOMA, HD-NOMA, and HD-OMA, let $R_{i}^{FD}=\frac{1}{3}R_{i}^{HD-OMA}$ and $R_{i}^{HD}=\frac{1}{2}R_{i}^{HD-OMA}$ for $ i \in \{1, gs\}$.

\subsection{Mathematical Preliminaries}

For $N$ communicating nodes, let $X_i, 0 \leq i \leq N$ be the Rician shadowed distributed random variable (RV) with Rician $K$ factor $K_i$ and shadowing severity parameter $m_i$. Furthermore, let $X_0$ be the desired node and the remaining $N$ nodes be the interferers. Then, the cumulative distribution function (CDF) expansion of $X_0$ is written in the following lemma.

\begin{lemma} \label{cdf_exp}
The cumulative distribution function (CDF) expansion $\alpha\big(n,\overline{P}_0,K_0,m_0,\gamma\big)$ of the RV $X_0$ is \cite{tan2018ricianShad}:
\begin{eqnarray} 
\alpha\big(n,\overline{P}_0,K_0,m_0,\gamma\big) & = & \sum_{i=0}^n (-1)^{n-i} \bigg(\frac{m_0}{K_0+m_0}\bigg)^{m_0} \frac{(m_0)_i}{\Gamma^2(i+1)} \nonumber \\
& & \hspace{-1.5cm} \times \bigg(\frac{K_0}{K_0+m_0}\bigg)^{i} \bigg(\frac{1+K_0}{\overline{P}_0}\bigg)^{n+1} \frac{\gamma^{n+1}}{(n-i)!(n+1)}_,
\end{eqnarray}
\end{lemma}
where $(m_0)_i=\frac{\Gamma(m_0+i)}{\Gamma(m_0)}$ is the Pochhammer symbol \cite{chun2017comprehensive}, $\overline{P}_i$ is the variance of $X_i$, and $\gamma$ is a threshold value.

Defining the signal-to-interference-plus-noise ratio (SINR) as $\frac{X_0}{1+\sum_{i=1}^{N}X_i}$, the outage probability is presented in the following Lemma.

\begin{lemma} \label{lemma_P_out}
For an outage event $\mathcal{O}$ at an arbitrary receiver, the outage probability is \cite[eq. (12)]{rached2017unified}:
\begin{eqnarray} 
Pr\big(\mathcal{O}\big) & \approx & \sum_{n=0}^{K_{tr}} \sum_{l_1+\ldots+l_{N+1}=n+1} \alpha\big(n,\overline{P}_0,K_{0},m_0,\gamma\big) \nonumber\\
& & \hspace{2cm} \times \binom{n+1}{l_1,\ldots, l_{N+1}} \prod_{j=1}^{N}  E\{X_{j}^{l_j}\}_,
\end{eqnarray}
\end{lemma}
where $\mathcal{O}=\Big\{ X_{0}, X_{i} : R \geq \log_{2}\Big(1 + \frac{X_0}{1+\sum_{i=1}^{N}X_i}\Big)\Big\}$, $R$ is the transmission rate, $K_{tr}$ is the truncation order and $E\{\bullet\}$ is the statistical expectation. For a Rician shadowed distributed RV, the $l_j^{th}$ moment of $X_{j}$ \cite[eq. (10)]{chun2017comprehensive} is
\begin{eqnarray} 
E\{X_{j}^{l_j}\} & = & \bigg(\frac{\overline{P}_j}{1+K_j}\bigg)^{l_j} \Gamma(1+l_j) \bigg(\frac{m_j}{K_j+m_j}\bigg)^{m_j-1-l_j} \nonumber\\
& & \hspace{1.7cm} \times {}_2{F_1}\bigg(1-m_j,1+l_j;1;-\frac{K_{j}}{m_j}\bigg)_,
\end{eqnarray}
where $_2F_1(\bullet)$ is the Gauss hypergeometric function \cite{kumar2015approximate}. 

\subsection{Full-Duplex NOMA Outage Probability}

From Lemma \ref{lemma_P_out}, one obtains the outage probability expressions at the FD-GS, UAV-2, and UAV-3.

\subsubsection{FD-GS}
Let the instantaneous received power of the SOI at the FD-GS be $X_{1,g}=P_{1,g}|h_{1,g}|^2$, where $P_{1,g}=\frac{P_t}{(d_{1,g})^n}$ and $X_{1,g}$ is a Rician shadowed distributed RV with Rician $K$ factor $K_{1,g}$ and shadowing parameter $m_{1,g}$. The instantaneous received power of the residual SI components are defined as $Y_{si,1}=P_{si}\gamma_{\phi}^2|h_{si}|^2$ and $Y_{si,2}=P_{si}\epsilon|\widetilde{h}_{si}|^2$, where $P_{si}=P_{t}$. The variable $Y_{si,1}$ is a Rician shadowed distributed RV, with Rician $K$ factor $K_{si,1}$ and shadowing parameter $m_{si,1}$, while $Y_{si,2}$ is an exponentially distributed RV.

Let the FD-GS outage event be defined as $\mathcal{O}_{gs}^{FD-NOMA} = \Big\{ h_{1,g}, h_{si} : R_{1}^{FD} \geq \log_{2}\Big(1 + \frac{X_{1,g}}{Y_{si,1} + Y_{si,2} + 1}\Big)\Big\}$, with threshold $\gamma_{gs}^{FD} = 2^{R_{1}^{FD}}-1$. Then, the outage probability $Pr\big(\mathcal{O}_{gs}^{FD-NOMA}\big)$ at the FD-GS is presented in the following theorem:

\begin{theorem} \label{theorem_P_out_fd_gs}
The outage probability at the FD-GS is
\begin{eqnarray} \label{P_out_fd_gs}
Pr\big(\mathcal{O}_{gs}^{FD-NOMA}) \hspace{-0.25cm} & \approx & \hspace{-0.25cm} \sum_{i=0}^{K_{tr}} \sum_{l_1+l_2+l_3=i+1} \alpha\big(i,P_{1,g},K_{1,g},m_{1,g},\gamma_{gs}^{FD}\big) \nonumber\\ 
& & \hspace{1.2cm} \times \binom{i+1}{l_1, l_2, l_3}  E\{Y_{si,1}^{l_1}\} E\{Y_{si,2}^{l_2}\}
\end{eqnarray}
\end{theorem}
\begin{IEEEproof}
Theorem \ref{theorem_P_out_fd_gs} is obtained from Lemma \ref{lemma_P_out} by substituting $X_0=X_{1,g}$ and treating $Y_{si,1}$ and $Y_{si,2}$ as interferers.
\end{IEEEproof}

The outage probability expression in (\ref{P_out_fd_gs}) enables residual SI to be taken into consideration at the FD-GS.

\subsubsection{Downlink UAVs}

As UAV-2 is nearer to the FD-GS than UAV-3, we consider an imperfect SIC detector and an II detector at UAV-2 and UAV-3, respectively. Let the instantaneous received power of the SOI at UAV-$i, i \in \{2,3\}$ be $X_{g,i}=P_{g,i}|h_{g,i}|^2$, where $P_{g,i}=\frac{P_t}{\sigma_i^2(d_{g,i})^n}$. Similarly, let $Y_{1,i}=P_{1,i}|h_{1,i}|^2, i \in \{2,3\}$ be the instantaneous received power of the inter-UAV interference due to the principles of FD-NOMA transmissions, where $P_{1,i}=\frac{P_t}{\sigma_i^2(d_{1,i})^n}$. The variables, $X_{g,i}$ and $Y_{1,i}$, are Rician shadowed distributed RVs with respective Rician $K$ factors $K_{g,i}$ and $K_{1,i}$, and shadowing parameters, $m_{g,i}$ and $m_{1,i}$.

At UAV-$i$, the outage event is defined as $\mathcal{O}_{i}^{FD-NOMA} = \Big\{ h_{g,i}, h_{1,i} : \frac{X_{g,i}}{Y_{1,i} + 1} \leq \gamma_{i}^{FD*}\Big\}$, where $\gamma_{2}^{FD*} = \frac{\gamma_{2}^{FD}}{a_{gs,2} - (1-a_{gs,2})\beta\gamma_{2}^{FD}}$, $\gamma_{3}^{FD*} = \frac{\gamma_{3}^{FD}}{a_{gs,3} - (1-a_{gs,3})\gamma_{3}^{FD}}$, $\gamma_{i}^{FD} = 2^{R_{gs}^{FD}}-1$, and $0 \leq \beta \leq 1$ denotes the strength of the residual interference from UAV-3 after SIC \cite{wang2017sir}. In the following theorem, the outage probability expression for UAV-$i$ is presented.

\begin{theorem} \label{theorem_P_out_down_fd_uav}
The FD-NOMA outage probability at UAV-$i$ for $i \in \{2,3\}$ is
\begin{eqnarray} \label{P_out_down_fd_uav}
& & \hspace{-1.1cm} Pr\big(\mathcal{O}_{i}^{FD-NOMA}) \nonumber \\
 & \hspace{-0.4cm} \approx & \hspace{-0.3cm} \sum_{q=0}^{K_{tr}} \sum_{l=0}^{q+1} \alpha\big(q,P_{g,i},K_{g,i},m_{g,i},\gamma_{i}^{FD*}\big) \binom{q+1}{l} E\{Y_{1,i}^{l}\}_,
\end{eqnarray}
\end{theorem}
\begin{IEEEproof}
Substituting $X_0=X_{g,i}$, $X_1=\widetilde{X}_{g,i}$, and $X_2=Y_{1,i}$ into Lemma \ref{lemma_P_out} yields (\ref{P_out_down_fd_uav}).
\end{IEEEproof}

\subsection{Benchmark Schemes}
In this work, HD-NOMA and HD-OMA are used as benchmark schemes against FD-NOMA.

\subsubsection{Half-Duplex NOMA Outage Probability}

Under HD-NOMA, SI at the GS and inter-UAV interference due to FD mode at the downlink UAVs are non-existent. At the HD-GS, the outage event is defined as $\mathcal{O}_{gs}^{HD-NOMA} = \big\{ h_{1,g} : R_{1}^{HD} \geq \log_{2}\big(1 + X_{1,g}\big)\big\}$. The outage probability $Pr\big(\mathcal{O}_{gs}^{HD-NOMA}\big)$ is presented in the following theorem.

\begin{theorem} \label{theorem_P_out_hd_gs}
The outage probability at the HD-GS is
\begin{eqnarray} \label{P_out_hd_gs}
Pr\big(\mathcal{O}_{gs}^{HD-NOMA}) & \approx & \sum_{i=0}^{K_{tr}} \alpha\big(i,P_{1,g},K_{1,g},m_{1,g},\gamma_{gs}^{HD}\big)_,
\end{eqnarray}
where $\gamma_{gs}^{HD} = 2^{R_{1}^{HD}}-1$.
\end{theorem}
\begin{IEEEproof}
From Lemma \ref{cdf_exp}, expanding the CDF of $X_{1,g}$ and treating $\gamma = \gamma_{gs}^{HD}$ yields (\ref{P_out_hd_gs}).
\end{IEEEproof}

The outage event at downlink UAV-$i, i \in \{2,3\}$ for HD-NOMA is defined as $\mathcal{O}_{i}^{HD-NOMA} = \Big\{ h_{g,i}, h_{1,i} : X_{g,i}\leq \gamma_{i}^{HD*} \Big\}$, where $\gamma_{2}^{HD*} = \frac{\gamma_{2}^{HD}}{a_{gs,2} - (1-a_{gs,2})\beta\gamma_{2}^{HD}}$, $\gamma_{3}^{HD*} = \frac{\gamma_{3}^{HD}}{a_{gs,3} - (1-a_{gs,3})\gamma_{3}^{HD}}$, and $\gamma_{i}^{HD} = 2^{R_{gs}^{HD}}-1$. Using the HD-NOMA outage event, the outage probability at UAV-$i, i \in \{2,3\}$ is obtained in the following theorem.

\begin{theorem}
The HD-NOMA outage probability expression at UAV-$i$ for $i \in \{2,3\}$ is:
\begin{eqnarray} \label{P_out_down_hd_uav}
Pr\big(\mathcal{O}_{i}^{HD-NOMA}) & \approx & \sum_{q=0}^{K_{tr}} \alpha\big(q,P_{g,i},K_{g,i},m_{g,i},\gamma_{i}^{HD*}\big)_	,
\end{eqnarray}
\end{theorem}
\begin{IEEEproof}
Substituting $X_0=X_{g,i}$ and $X_1=\widetilde{X}_{g,i}$ into Lemma \ref{lemma_P_out} yields (\ref{P_out_down_hd_uav}).
\end{IEEEproof}

\subsubsection{Half-Duplex OMA Outage Probability}

At the GS, the HD-OMA outage event is defined as $\mathcal{O}_{gs}^{HD-OMA} = \big\{ h_{1,g} : R_{1}^{HD-OMA} \geq \log_{2}\big(1 + X_{1,g}\big)\big\}$. Following the steps in (\ref{P_out_hd_gs}) yields the following outage probability at the GS:
\begin{eqnarray} \label{P_out_hd_oma_gs}
\hspace{-0.2cm} Pr\big(\mathcal{O}_{gs}^{HD-OMA}) \hspace{-0.2cm} & \approx & \hspace{-0.2cm} \sum_{i=0}^{K_{tr}} \alpha\big(i,P_{1,g},K_{1,g},m_{1,g},\gamma_{gs}^{HD-OMA}\big)_,
\end{eqnarray}
where $\gamma_{gs}^{HD-OMA} = 2^{R_{1}^{HD-OMA}}-1$.

At the downlink UAVs, the HD-OMA outage event is defined as $\mathcal{O}_{i}^{HD-OMA} = \Big\{ h_{g,i} : R_{gs}^{HD-OMA} \geq \log_{2}\Big(1 + X_{g,i}\Big)\Big\}$ with outage probability at UAV-$i$ for $i \in \{2,3\}$ given as:
\begin{eqnarray}
\hspace{-0.2cm} Pr\big(\mathcal{O}_{i}^{HD-OMA}) \hspace{-0.2cm} & \approx & \hspace{-0.2cm} \sum_{q=0}^{K_{tr}} \alpha\big(q,P_{g,i},K_{g,i}m,_{g,i},\gamma_{i}^{HD-OMA}\big)_,
\end{eqnarray}
where $\gamma_{i}^{HD-OMA} = 2^{R_{gs}^{HD-OMA}}-1$.

\section{Numerical Results and Simulation Studies} \label{sec_num_res}

In this section, the outage probabilities at the GS and the downlink UAVs are presented together with Monte Carlo simulations conducted with $10^{6}$ samples. To model the suburban environment, a Rician $K$ factor of $10$ is assumed for all links \cite{matolak2017air_suburban}. For the shadowing parameters, we let $m_{1,g}=m_{si,1}=m_{g,3}=m_{1,3}=10$ and $m_{g,2}=m_{1,2}=3$. Thus, the effect of shadowing is more severe at UAV-2 than at the GS and UAV-3. We also choose the noise variance at GS and the downlink UAVs to be set at $-131dBm$ \cite{ITU2011}. Finally, we let $R_{i}^{HD-OMA} = 0.2$ for $i \in \{1, gs\}$, $a_{gs,2} = 0.5$, $\beta = 0.1$, $\gamma_{\phi}^2 = -140 dBm$, $\epsilon = 0.1$, $d_{1,g} = 3$, $d_{g,2} = 2$, $d_{g,3} = 3$, and $n = 2$.

\begin{figure} []
\centering
\includegraphics [width=0.95\columnwidth]{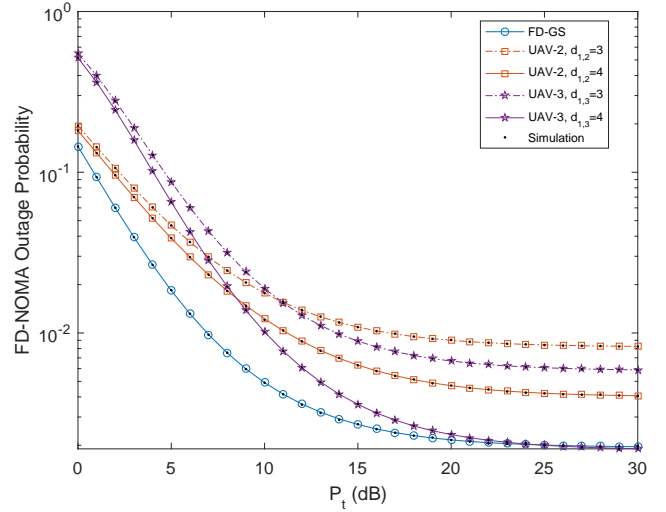} 
\caption{FD-NOMA outage probability comparison at the GS and the downlink UAVs.}
\label{fig:fd_noma_outage}
\end{figure}


The FD-NOMA outage probabilities at the GS and the downlink UAVs are plotted in Fig. \ref{fig:fd_noma_outage}. At low $P_t$ regimes, it is observed that the downlink UAVs form the bottleneck of performance in FD-NOMA. At UAV-2, more severe shadowing is experienced than at UAV-3 and at the FD-GS. However, UAV-2 is in closer proximity to the FD-GS than UAV-3. As a result, inter-UAV interference due to power domain NOMA from UAV-3 and uplink interference from UAV-1 is less detrimental than the impact of shadowing. Thus, an outage probability that is lower than UAV-3 is observed at UAV-2 at low $P_t$ regimes, e.g., $P_t < 10dB$. 

At high $P_t$ regimes, it can be seen from Fig. \ref{fig:fd_noma_outage} that the FD-GS and the downlink UAVs are interference-limited. In particular, an error floor is observed at the FD-GS due to the presence of SI. Thus, the FD-GS is SI-limited, i.e., limited by SI, as $P_t \to \infty$. In contrast, UAV-2 is limited by inter-UAV interference from UAV-1 and UAV-3 as $P_t \to \infty$. The former is due to uplink interference being ignored at UAV-2 while the latter is due to the imperfect SIC detector. Likewise, similar trends are also observed for UAV-3 due to inter-UAV interference from UAV-1 and UAV-2.

Furthermore, it is observed that UAV-3 achieves lower outage probability than UAV-2 at high $P_t$ regimes, despite the adoption of the SIC detector at the latter. Such an observation shows that the residual interference due to the imperfect SIC detector is more detrimental than inter-UAV interference from UAV-1 at high $P_t$ regimes. Thus, it is evident that sophisticated detectors will be needed at the downlink UAVs to better manage inter-UAV interference.

\begin{figure} []
\centering
\includegraphics [width=0.95\columnwidth]{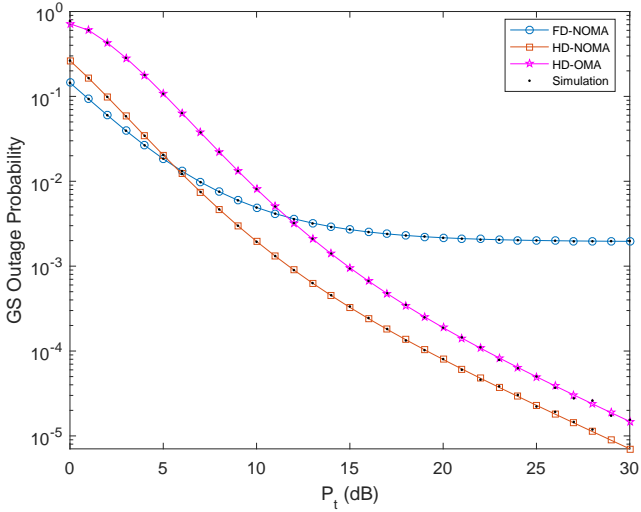} 
\caption{Outage probability comparison of FD-NOMA, HD-NOMA, and HD-OMA at the GS.}
\label{fig:gs_outage}
\end{figure}

In Fig. \ref{fig:gs_outage}, the GS outage probability is plotted. At low $P_t$ regimes, the GS attains lower outage probability when operating in FD-NOMA mode than in HD-NOMA and HD-OMA modes. However, at high $P_t$ regimes, the HD-NOMA and HD-OMA modes attains lower outage probability than FD-NOMA due to the absence of SI. Additionally, it is observed that HD-NOMA attains lower outage probability than HD-OMA due to a lower threshold, i.e., $\gamma_{gs}^{HD}<\gamma_{gs}^{HD-OMA}$.

\begin{figure} []
\centering
\includegraphics [width=0.95\columnwidth]{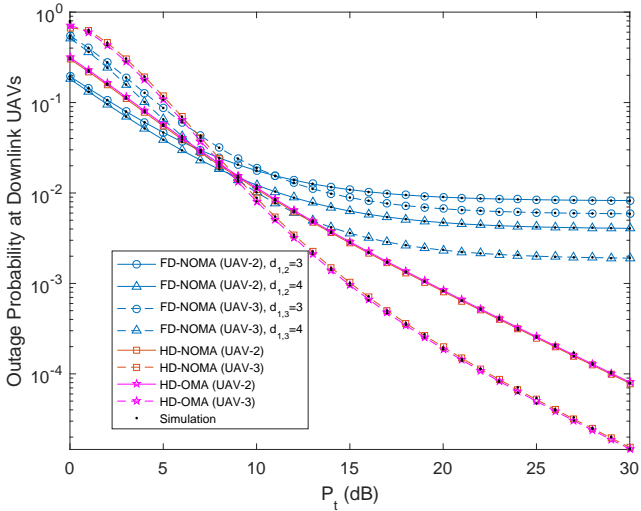} 
\caption{Outage probability comparison of FD-NOMA, HD-NOMA, and HD-OMA at the downlink UAVs.}
\label{fig:downlink_outage}
\end{figure}

The outage probabilities at the downlink UAVs are plotted in Fig. \ref{fig:downlink_outage}. At the downlink UAVs, FD-NOMA attains lower outage probability than HD-NOMA and HD-OMA at low $P_t$ regimes. Furthermore, both HD-NOMA and HD-OMA attain lower outage probability at UAV-2 than at UAV-3 due to a shorter distance between the GS and UAV-2, i.e., $d_{g,2}<d_{g,3}$.

At high $P_t$ regimes, FD-NOMA attains higher outage probability than HD-NOMA. As $P_t$ increases, both FD-NOMA and HD-NOMA become interference-limited. For FD-NOMA, the downlink UAVs are limited by inter-UAV interference due to power-domain NOMA and uplink interference from UAV-1. Additionally, the severity of shadowing at UAV-2, along with residual interference due to the imperfect SIC detector, leads to UAV-2 exhibiting higher outage probability than UAV-3 at high $P_t$ regimes. In contrast, uplink interference from UAV-1 is absent in HD-NOMA and HD-OMA. Thus, both HD-NOMA and HD-OMA are interference-free. Additionally, for HD-OMA, the severity of shadowing leads to UAV-2 attaining higher outage probability than UAV-3 at high $P_t$ regimes. From Fig. \ref{fig:downlink_outage}, it is clear that more advanced detector will be needed to manage the combined effects of fading, shadowing and interference at the downlink UAVs.

\section{Conclusion} \label{sec_conclusion}
In this work, the feasibility of FD-NOMA is investigated to address spectrum scarcity in UAV communications. To this end, closed-form outage probability expressions for FD-NOMA, HD-NOMA, and HD-OMA are presented for UAV communications over Rician shadowed fading channels. Performance analysis at the FD-GS and downlink UAVs showed FD-NOMA attaining lower outage probability than HD-NOMA and HD-OMA at low transmit power regimes. In contrast, residual SI and inter-UAV interference at the FD-GS and downlink UAVs causes FD-NOMA to be interference-limited at high transmit power regimes. At the system level, outage probability of FD-NOMA is constrained by the downlink UAVs. It is also shown that when shadowing is severe at UAV-2, the imperfect SIC detector attains higher outage probability than the II detector at UAV-3 when operating at high $P_t$ regimes. For the case of HD-OMA, severe shadowing at UAV-2 causes outage probability to be higher than UAV-3 at high transmit power regimes. Evidently, robust detectors will have to be considered at the downlink UAVs to mitigate the combined effect of fading, shadowing, and inter-UAV interference. Work is in progress towards investigating advanced interference management techniques for FD-NOMA at the downlink UAVs.

\ifCLASSOPTIONcaptionsoff
  \newpage
\fi

\bibliographystyle{IEEEtran}
\bibliography{IEEEabrv,ref}

\end{document}